\documentclass[journal]{IEEEtran}

\usepackage{blindtext, graphicx}
\usepackage{graphicx}
\ifCLASSINFOpdf
\else
\fi
\usepackage[cmex10]{amsmath}
\usepackage[cmex10]{mathtools}
\usepackage{amsfonts}
\usepackage{algorithmic}
\usepackage{caption}
\usepackage{bbm}
\usepackage{mathrsfs}
\usepackage{tabularx}

\graphicspath{{Pictures/}}
\begin{document}

%
\title{Localized solar power prediction based on weather data from local history and global forecasts}

\author{\IEEEauthorblockN{Chaitanya Poolla\IEEEauthorrefmark{1},
Abraham K. Ishihara\IEEEauthorrefmark{2}}\\
\IEEEauthorblockA{ECE,
Carnegie Mellon University (SV)\\
Moffett Field, CA 94035\\
Email: \IEEEauthorrefmark{1}chaitanya.poolla@west.cmu.edu,
\IEEEauthorrefmark{2}abe.ishihara@west.cmu.edu}}
%
%

\maketitle


\pagenumbering{gobble}

\begin{abstract}
With the recent interest in net-zero sustainability for commercial buildings, integration of photovoltaic (PV) assets becomes even more important. This integration remains a challenge due to high solar variability and uncertainty in the prediction of PV output. Most existing methods predict PV output using either local power/weather history or global weather forecasts, thereby ignoring either the impending global phenomena or the relevant local characteristics, respectively. This work proposes to leverage weather data from both local weather history and global forecasts based on time series modeling with exogenous inputs. The proposed model results in eighteen hour ahead forecasts with a mean accuracy of $\approx$ 80\% and uses data from the National Ocean and Atmospheric Administration's (NOAA) High-Resolution Rapid Refresh (HRRR) model.
\end{abstract}
%
\IEEEpeerreviewmaketitle
\section{Introduction}
\label{sec:intro}
The power output from a PV array is known to depend on environmental variables such as irradiance, temperature (ambient and cell), wind velocity, relative humidity, air pressure, and sky conditions \cite{CHEN20112856} \cite{localfactorsNgyuen}. Previous works have shown that the lack of accurate information about these variables can affect the prediction error significantly \cite{Hiyama_1997}. Therefore, it is important to be able to forecast these fluctuations accurately. Using such forecast plus a model that maps the environmental states to the renewable output, we can forecast the renewable output. While there exist several forecast products from the National Oceanic and Atmospheric Administration (NOAA), the spatial resolution of each of these products are typically on the order of kilometers \cite{center1979national}. Resolution at such scales is inadequate for purposes of localized predictions over smaller spatial scales. Therefore, a blend of local weather history along with global weather forecasts is critical to improve local weather forecast accuracy.
%
%

In this work, we are concerned with the problem of weather forecast-based solar power prediction. Methods proposed in the literature to forecast solar power are either based on time-series power data, global weather forecast data, or local weather measurement data. In models such as those in Pedro et al. \cite{pedro2012assessment}, time-series power data with no exogenous inputs are used for short-term forecasting with a horizon of up to two hours into the future. However, using only past power data for forecasting does not directly integrate globally induced weather-related changes into the forecasting model. In \cite{bacher2009online} \cite{yona2007application}, both time-series power data and global weather forecasts based on mesoscale models were used for forecasting. Bacher et al. \cite{bacher2009online} conclude that for horizons up to less than six-hours, solar power is an important variable for prediction whereas over longer forecast horizons greater than nineteen hours ahead, only weather input was found adequate for prediction. However, using only global weather forecast data does not incorporate local characteristics such as shadows due to trees, buildings, or birds into the weather forecast model. On the contrary, there are also approaches that use local power and weather history for prediction as described by Chen et al. \cite{chen2011online}. However, using only local data does not allow the model to account for impending global phenomena. As a result it is imperative to develop models that incorporate local and global characteristics to improve prediction accuracy.

Consequently, this work proposes a two-step approach to predict local power data based on past local weather data and global weather forecast data. In the first step, a local weather prediction problem is solved by employing a time-series model with past local weather data and global weather forecast data as inputs. The local weather prediction results from the first step are used as inputs in the second step to predict the solar power output based on an existing weather-to-power map \cite{wan2015photovoltaic}. In this work, the solar irradiance, outside air temperature, and wind speed are the weather variables considered in the weather prediction problem. Out of these variables, the solar irradiance and outside air temperature are considered in the power prediction problem. The remainder of this paper is structured as follows. Notation is presented in Section \ref{sec:notations} and the weather data is described in \ref{sec:noaa_weather}. The two-step model is described in Section \ref{sec:model} and is followed by results and discussion in Section \ref{sec:results_and_discussion}. Concluding remarks are presented in Section \ref{sec:conclusion}.
\section{Notation}
\label{sec:notations}
Let the elements of the sequence $\{t_i\}_{i=-\infty}^\infty$ represent time instances in the past ($i<0$), the present ($i=0$), and the future ($i>0$). The weather variables such as solar irradiance, outside air temperature, and wind speed may either be measured or HRRR-forecasted or predicted. Let these measured variables, j-hour ahead forecast variables, and variables predicted at the time instant $t_i$ be represented by the ordered triples $(I^M_i,T^M_i,W^M_i)$, $(I^F_{i,j},T^F_{i,j},W^F_{i,j})$, and $(\hat{I}_i,\hat{T}_i,\hat{W}_i)$, respectively. Further, at the time instant $t_i$, let the measured or analytically determined power be represented by $P_i$ and let the predicted power be represented by $\hat{P}_i$.
%

\section{NOAA weather data}
\label{sec:noaa_weather}
The NOAA's National Center for Environmental Prediction (NCEP) provides several forecast products differing in forecast horizon, spatio-temporal resolution, update frequency, forecast variables, and forecast method \cite{noaa_ncep_prod}. In this work, we consider the High Resolution Rapid Refresh (HRRR) product which offers weather data at a spatial resolution of three kilometers \cite{noaa_hrrr}. The temporal resolution of the HRRR data is either one hour or fifteen minute depending on whether an hourly or a subhourly product is used \cite{nomads_hrrr}. An archive of the HRRR hourly data is available from the University of Utah MesoWest HRRR archive \cite{BLAYLOCK201743}. This data is temporally organized into twenty four model cycles each reflecting an hourly update during the day. Within each model cycle, forecast files are provided for up to eighteen hours ahead at a temporal resolution of one hour. Each model cycle also contains a zero hour ahead forecast, which is an assimilation of observations from several primary sources. For this work, we consider this assimilated value as the reference or measured value of the corresponding weather variable at the location of interest.

In line with the available HRRR archive data, we let the time instances $t_i \forall i \in \mathbb{Z}$ to be an hour apart from each other so that $\Delta t = (t_{i+1}-t_i)$ = 3600 seconds. Accordingly, during the present hour $t_i (i=0)$, the weather measurements $(I^M_0,T^M_0,W^M_0)$ and forecasts up to eighteen hours ahead $(I^F_{i,j},T^F_{i,j},W^F_{i,j}) \forall i \in \{1,\cdots,18\}, 1\leq j\leq 18$ are well-defined. In this work, the forecast horizon is set to eighteen hours ahead while noting similar analyses may be performed for shorter forecast horizons with the HRRR forecast data. Henceforth, the eighteen hour ahead forecasts $(I^F_{i,18},T^F_{i,18},W^F_{i,18})$ for the time instant $t_i$ will be represented succinctly as $(I^F_i,T^F_i,W^F_i)$.

\section{Model}
\label{sec:model}
We employ the following two-step model to predict the eighteen hour ahead solar power output based on the measured weather data and the eighteen hour ahead forecast weather data.

\subsection{Step 1: Weather Prediction}
\label{ssec:step_1}
In the weather prediction problem we seek to predict the eighteen hour ahead weather data $(\hat{I}_i,\hat{T}_i,\hat{W}_i)$ at the instant $t_i$ based on the 24-hour behind measured data $(I^M_{i-24},T^M_{i-24},W^M_{i-24})$ and 18-hour ahead forecast data $(I^F_i,T^F_i,W^F_i)$. We resort to time-series modeling and develop an autoregressive model ARX(1,1) with past weather measurements and exogenous weather forecasts as inputs. A comparison to the reference model (AR) without the exogenous forecast input will then illustrate the utility of weather forecasts in the otherwise local history-based predictions. This comparison is discussed Section \ref{sec:results_and_discussion}. Accordingly, let the ARX(1,1) model be represented by:
\begin{equation}
\label{eqn:arx_model_step1}
\hat{X}_i = \alpha X^M_{i-24} + \beta X^F_i + \gamma + \epsilon_i
\end{equation}
where, $X$ represents each of the weather variables, $(\alpha,\beta)$ represent the model coefficients, and $(\gamma,\epsilon_i)$ represent the bias and the error term. Correspondingly, the reference model can be degenerated from the above ARX(1,1) model as shown in equation \ref{eqn:ar_model_step1}.
\begin{equation}
\label{eqn:ar_model_step1}
\hat{\underline{X}}_i = \underline{\alpha} X^M_{i-18} + \underline{\gamma} + \underline{\epsilon_i}
\end{equation}
where, $\hat{\underline{X}}_i$ represents the weather prediction from the reference model and $(\underline{\alpha},\underline{\gamma},\underline{\epsilon_i})$ represent the model parameters along with the error term. Both these models are trained over the HRRR dataset to learn the corresponding model parameters. The trained model is used to predict local weather variables in the future.
\subsection{Step 2: Solar Power Prediction}
\label{ssec:step_2}
Once the weather prediction model is developed, a weather-to-power mapping is used to translate the weather predictions $(\hat{I}_i,\hat{T}_i,\hat{W}_i)$ into solar power predictions $\hat{P}_i$. While accurate physics-based or sophisticated data-based mappings can be employed in principle \cite{sera2007pv} \cite{poolla2014neural}, we resort to a linear model to demonstrate the concept using the relation from \cite{wan2015photovoltaic} as shown in the equation below:
\begin{equation}
\label{eqn:model_step2}
\hat{P}_i = \eta S \hat{I}_i [1-0.05(\hat{T}_i-298.15)]
\end{equation}
where, $\eta$ represents the panel efficiency, $S$ represents the panel area, and $298.15$ is the temperature under standard conditions. In this work, the usable area for a medium office building $S\approx1660 sqm.$ and a solar panel efficiency $\eta=16\%$ are considered as stated in Davidson et al. \cite{davidson2015nationwide}.
\section{Results and Discussion}
\label{sec:results_and_discussion}
\begin{figure}[h]
    \centering
    \includegraphics[width=0.5\textwidth]{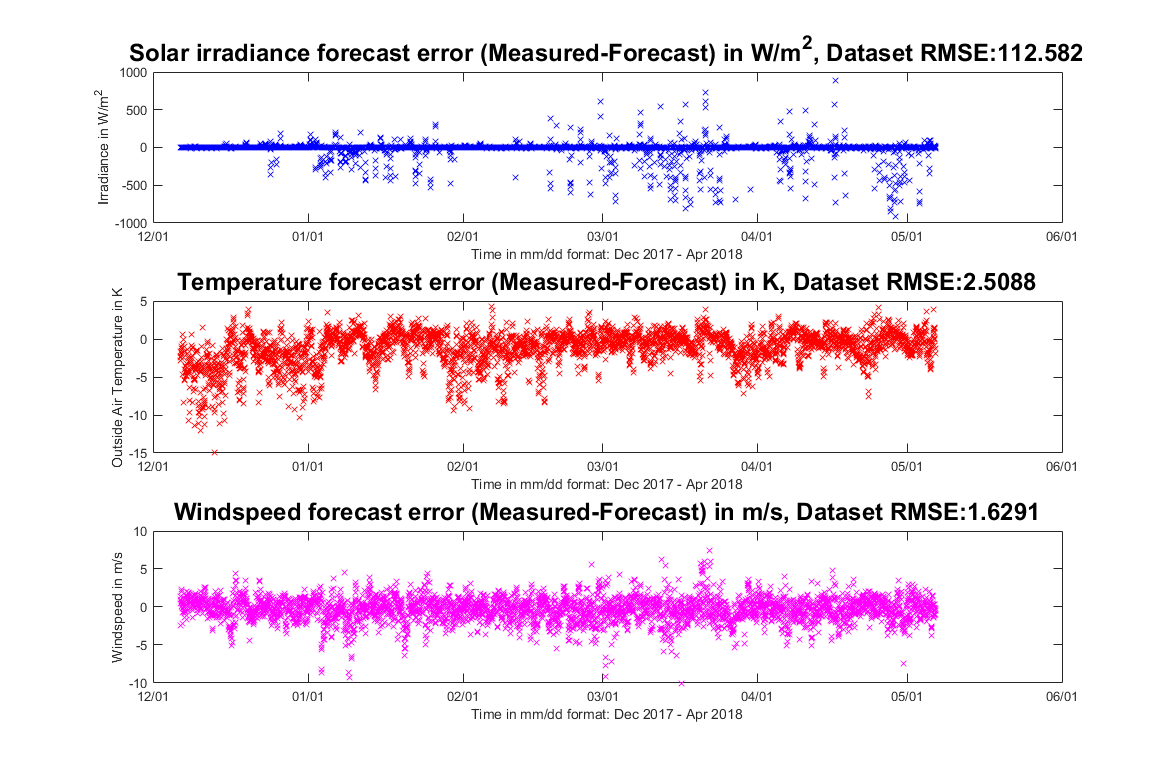}
    \caption{Forecast Error Characteristics}
    \label{fig:scatter_fcast_error}
\end{figure}

\begin{figure}[h]
    \centering
    \includegraphics[width=0.5\textwidth]{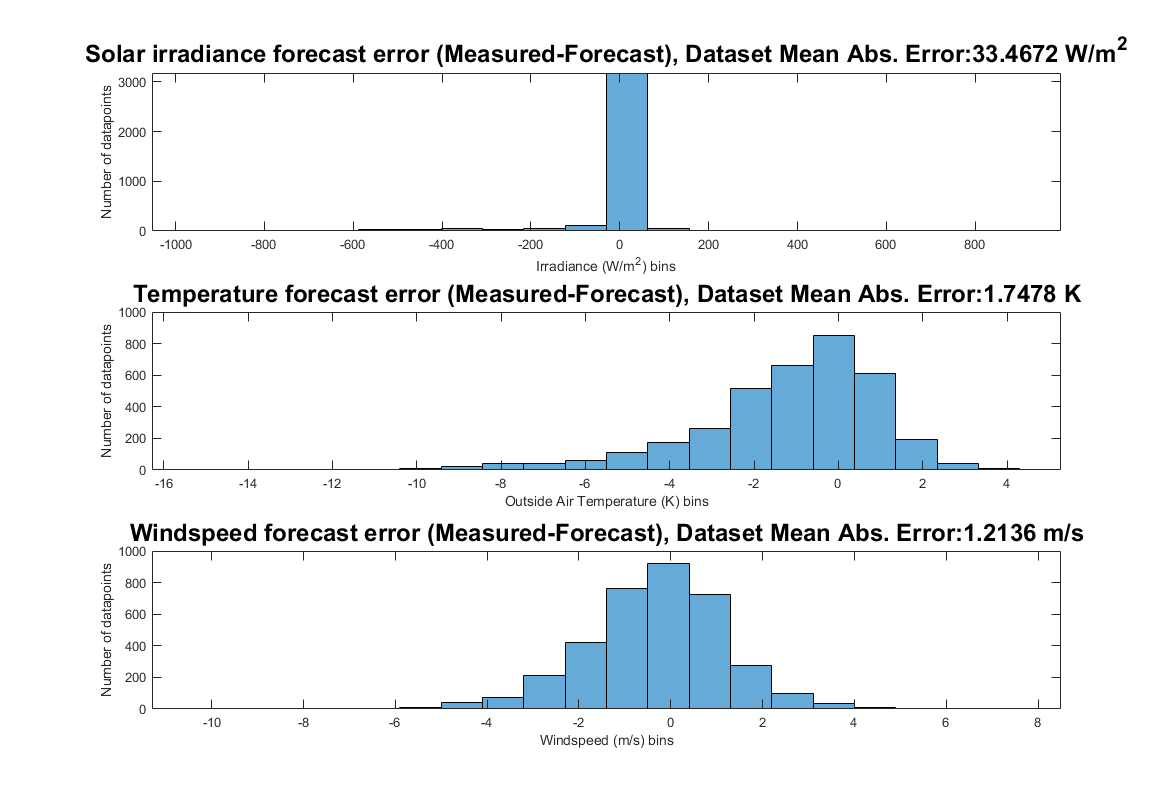}
    \caption{Forecast Error Histogram}
    \label{fig:hist_fcast_error}
\end{figure}

\begin{figure}[h]
    \centering
    \includegraphics[width=0.5\textwidth]{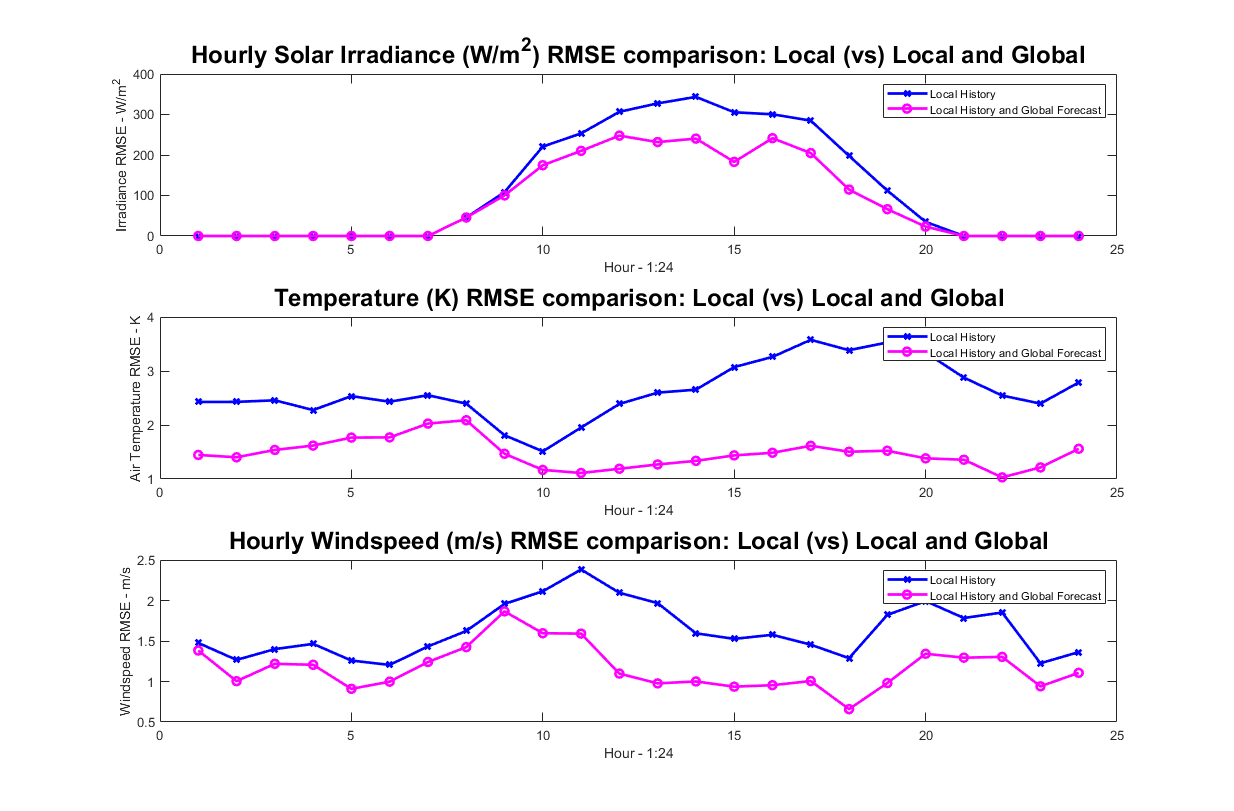}
    \caption{Comparison of exclusive local history model and local history with global forecast model}
    \label{fig:hourly_rmse_plot}
\end{figure}

\begin{figure}[h]
    \centering
    \includegraphics[width=0.5\textwidth]{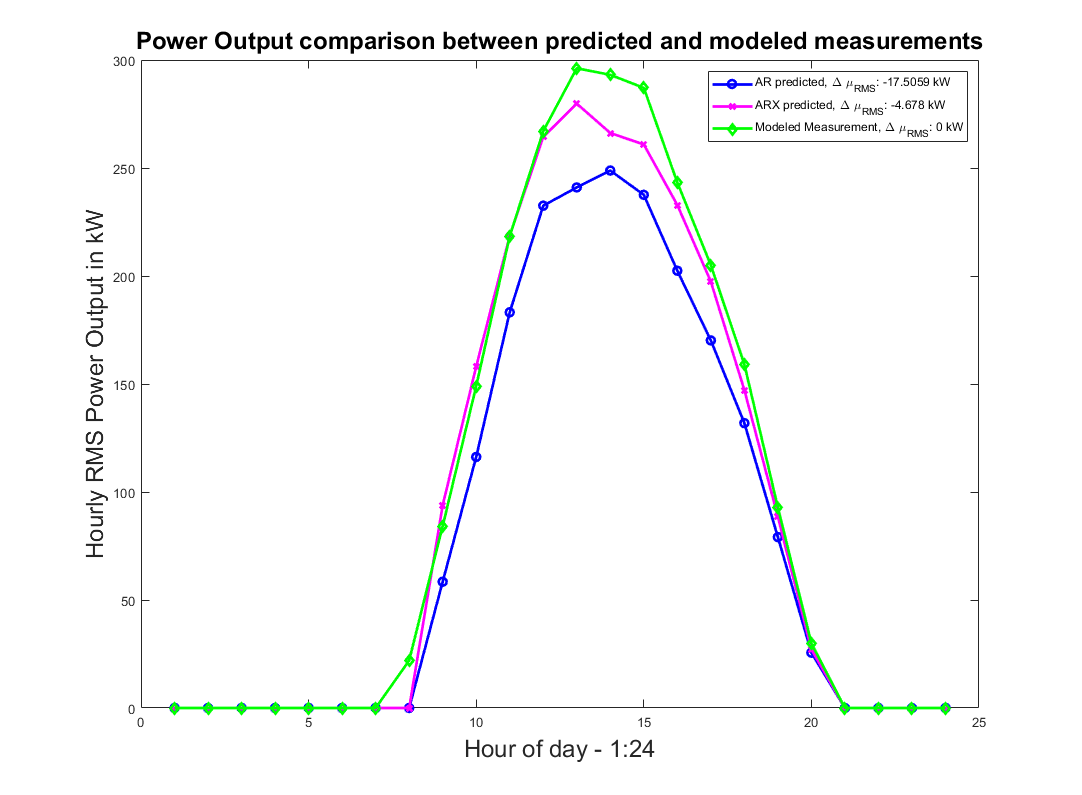}
    \caption{Comparison of AR and ARX power predictions against the weather-to-power-based measured power output}
    \label{fig:hourly_power_comparison}
\end{figure}

The dataset used in this work consisted of weather variables such as solar irradiance, temperature, and windspeed spanning over six months from Dec 2017 - May 2018 at Moffett Field, California. The error characteristics of the forecast against the measurements are shown in the Figure \ref{fig:scatter_fcast_error} and in the histogram \ref{fig:hist_fcast_error}. The mean error metrics comparing the forecast dataset against the zero-hour ahead measurements are summarized in the Table \ref{table:hrrr_error_table}.
\begin{center}
	\captionof{table}{Error metrics from weather datasets} 
    \begin{tabular}{| l | p{1.5cm} | p{1.5cm} | p{1.5cm} |}
    \hline
    Metric\textbackslash Variable & Irradiance $(W/m^2)$ & Air Temp $(K)$ & Wind $(m/s)$ \\ \hline
    RMSE & 112.58 & 2.51 & 1.63 \\ \hline
    MAE & 33.47 & 1.75 & 1.21 \\ \hline
    \end{tabular}
    \label{table:hrrr_error_table}
\end{center}
For the weather modeling, the data points for each weather variable were classified into the hour of the day and the corresponding hourly model parameters were learnt based on the functional forms specified in equations \ref{eqn:arx_model_step1} and \ref{eqn:ar_model_step1}. Accordingly, twenty four hourly models were learnt each with and without exogenous inputs for all the weather variables of interest. The hourly datasets were split in the ratio 3:1 for training and testing purposes. The parameters of the models were estimated using the MATLAB System Identification Toolbox and are presented for ARX hourly model in the table \ref{table:model_coeff}. 
\begin{table}[htb]
\caption{Estimated Parameters (Mean) for the ARX(1,1) model $(\alpha_h,\beta_h,\sigma(\epsilon_h))$}
\label{1234}
\resizebox{\columnwidth}{!}{%
\begin{tabular}{c*{5}{>{$}c<{$}}}
\text{Hour (h)}    & \text{Irrad. Mdl.}        & \text{Air Temp. Mdl.}       & \text{Windspeed Mdl.}       \\
1  & (0,0,0) & (0.46,0.53,2.11) & (0.10,0.75,1.41) \\
2  & (0,0,0) & (0.39,0.60,2.00) & (0.25,0.59,1.22) \\
3  & (0,0,0) & (0.44,0.55,2.17) & (0.35,0.48,0.97) \\
4  & (0,0,0) & (0.51,0.49,2.29) & (0.35,0.50,1.18) \\
5  & (0,0,0) & (0.46,0.54,2.11) & (0.30,0.55,1.28) \\
6  & (0,0,0) & (0.42,0.57,2.31) & (0.24,0.55,1.31) \\
7  & (0,0,0) & (0.49,0.50,2.51) & (0.48,0.35,1.34) \\
8  & (0,0,0) & (0.41,0.58,2.51) & (0.23,0.67,1.42) \\
9  & (-0.08,0.99,15.03) & (0.48,0.52,2.46) & (0.22,0.68,1.33) \\
10  & (0.23,0.69,66.43) & (0.52,0.47,2.10) & (0.19,0.68,1.60) \\
11  & (0.14,0.76,108.03) & (0.45,0.55,1.65) & (0.09,0.74,1.51) \\
12  & (0.01,0.91,130.57) & (0.33,0.66,1.35) & (0.22,0.69,1.43) \\
13  & (0.15,0.77,195.29) & (0.33,0.67,1.37) & (0.23,0.70,1.48) \\
14  & (0.30,0.60,222.60) & (0.34,0.65,1.69) & (0.27,0.63,1.48) \\
15  & (0.14,0.74,200.10) & (0.30,0.69,1.65) & (0.18,0.73,1.42) \\
16  & (0.13,0.76,171.71) & (0.26,0.73,1.68) & (0.27,0.62,1.51) \\
17  & (0.20,0.71,136.80) & (0.20,0.80,1.56) & (0.22,0.67,1.48) \\
18  & (0.15,0.75,83.30) & (0.14,0.85,1.36) & (0.25,0.64,1.53) \\
19  & (0.24,0.68,32.88) & (0.10,0.90,1.21) & (0.12,0.73,1.59) \\
20  & (0.21,0.60,2.16) & (0.15,0.85,1.31) & (0.33,0.50,1.35) \\
21  & (0,0,0) & (0.33,0.66,1.44) & (0.32,0.57,1.51) \\
22  & (0,0,0) & (0.27,0.72,1.57) & (0.18,0.69,1.25) \\
23  & (0,0,0) & (0.35,0.64,1.63) & (0.12,0.73,1.15) \\
24  & (0,0,0) & (0.41,0.59,1.87) & (0.16,0.66,1.12) \\
\label{table:model_coeff}
\end{tabular}
}
\end{table}
Since the magnitudes of the local weather lagged predictor $X^M_{i-24}$ and the corresponding forecast predictor $X^F_i$ are comparable, the coefficient magnitudes from the above model indicate that the forecasts are matter more for the prediction compared to the local history. In other words, the forecast variable is found to be more important than local history for the eighteen hour ahead predictions which is analogous to Bacher et al's observations on the predominance of weather forecasts over local (power) history for predictions over longer horizons \cite{bacher2009online}. Plots describing the hour-to-hour RMSE variations from the AR and the ARX models are shown in Figure \ref{fig:hourly_rmse_plot}.
It was also found that the mean RMSE from the AR models was $(118.39 W/m^2, 2.63 K, 1.63 m/s)$ and the mean RMSE from the ARX model was $(86.84 W/m^2, 1.47 K, 1.17 m/s)$. Therefore, it is evident from both the trends and the average RMSE that the local history and global forecasts together provide a better forecast accuracy (RMSE) compared to predictions only based on the local history. The mean relative percentage accuracies (RMSE) for all the cases are summarized in the table \ref{table:perc_accuracy} below.
\begin{center}
	\captionof{table}{Percentage accuracy of different prediction models} 
    \begin{tabular}{| p{2.5cm} | p{1.5cm} | p{1.5cm} | p{1cm} |}
    \hline
    Model\textbackslash\ Prediction Accuracy & Irradiance $\%$ Accuracy & Air Temp $\%$ Accuracy & Wind $\%$ Accuracy \\ \hline
    AR (Local) & 73.42 & 80.67 & 52.51 \\ \hline
    HRRR (Global) & 79.31 & 87.10 & 56.58 \\ \hline
    ARX (Local and Global) & 80.07 & 88.55 & 64.38 \\ \hline
    \end{tabular}
    \label{table:perc_accuracy}
\end{center}
From these results, we observe that while the introduction of global forecasts to the local history-based AR prediction significantly improves the latter $(+6.65\%,+7.88\%,+11.87\%)$, the changes induced by adding local measurements to the global forecast-based predictions are relatively intangible in the irradiance and temperature predictions $(+0.76\%,+1.45\%,+7.80\%)$. This could be explained by the consideration that the local zero-hour ahead data is assimilated into the model for the forecasts. Therefore, there is not much additional information contained in the HRRR zero-hour ahead data given the global forecasts for forecasting the local weather. However, the outcome of adding local weather history to the global forecasts could be significant when the local history reflects measurements from onsite sensors capturing detailed local characteristics that may not be available from the assimilated zero-hour forecasts. During the second step, a weather-to-power mapping was used to predict the power output from a Medium Office building specified in Section \ref{ssec:step_2}. The results comparing the RMS predicted values from the AR and ARX models are shown in Figure \ref{fig:hourly_power_comparison}.
Assuming the validity of the power model, it can be observed that the predictions using both the local history and global forecasts resulted in a lower average RMS error ($|-4.68|\ kW$) compared to the average RMS error ($|-17.50|\ kW$) obtained by using on local history in the prediction model.
\section{Conclusion}
\label{sec:conclusion}
The problem of solar power prediction using local weather history and global forecast data was considered. The solution involved addressing a local weather prediction problem, whose output was fed into an existing weather-to-power model. The weather variables considered in this study were solar irradiance $(W/m^2)$, outside air temperature $(K)$, and windspeed  $(m/s)$. The weather data was obtained from the HRRR forecast archive wherein the zero-hour ahead forecast represented the local weather measurements and the forecasts up-to eighteen hours into the future represented the global weather forecasts. The local weather prediction model was constructed based on hourly autoregressive time-series functions fitted using $75\%$ of the dataset (Dec 2017 - Apr 2018) and tested against the rest (Apr 2018 - May 2018). Results indicate that using both the local history and global forecasts for prediction results in a higher mean RMS accuracy ($80.07\%$) in comparison to the predictions using local history alone ($73.42\%$). However, only marginal differences were visible in the irradiance and temperature predictions by adding zero-ahead history to the prediction mechanism which were otherwise obtained directly from the HRRR forecast. On a $265 kW$ rated solar array setup considered for medium office buildings, power output predictions using an existing weather-to-power mapping demonstrated a lower mean RMS error of $4.68 kW$ using both local history and global forecasts in comparison to $17.50 kW$ using the local history alone. Future studies can investigate the generalizability of the results across different forecasts, locations, timespans, and horizons using non-NOAA-based on-site sensor measurements for representing local weather data.

\bibliographystyle{IEEEtran}
\bibliography{allreferences}
\end{document}